# A Bayesian network approach for assessing the general resilience of road transportation systems: A systems perspective

Junqing TANG[1]*, Hans R. HEINIMANN[1] and Ke HAN[2]

[1] ETH Zurich, Future Resilient Systems, Singapore-ETH Centre, 1 CREATE Way, CREATE Tower, 138602 Singapore; E-mail: junqing.tang@frs.ethz.ch (Corresponding Author); E-mail: hans.heinimann@env.ethz.ch

[2] Department of Civil and Environmental Engineering, Imperial College London, London SW7 2BU, UK; E-mail: k.han@imperial.ac.uk

**ABSTRACT**

We proposed a Bayesian Network Model (BNM) based on function-oriented resilience framework and ontological interdependence among 10 system qualities to probabilistically assess the general resilience of the road transportation system in Beijing from 1997 to 2016. We tested the model with multi-source data collected from various sectors. The system qualities were examined by analysis of sensitivity and influence. The result shows that the general resilience of Beijing's road system exhibits a "V" shape in its trend, with the probability of being generally resilient between 50% and 70%, and at its minimum in 2006. There was a steep increase in such a probability since 2006. In addition, the general resilience of Beijing's road transportation system is most affected by its capabilities: (1) to rebuild its performance, (2) to be robust, (3) to adapt, (4) to change, and (5) to quickly repair damaged parts. The proposed BNM is a promising tool for multi-dimensional and systematic analysis, instead of finding a one-size-fits-all quantification criterion for the resilience.

## INTRODUCTION

Building resilience in critical infrastructure systems is becoming one of the most concerned priorities for urban governors and practitioners. In this connection, a resilient transportation system is not an exception and assessing the resilience of a city's transportation system is the cornerstone of the study. The methods on how to measure and, most importantly, how to interpret transportation resilience has been continuously developed with topological approaches and system-based approaches (Reggiani et al., 2015; Mattsson and Jenelius, 2015).

The mainstream of discussing transportation resilience relies on the application of the percolation theory. Most of the previous works study transportation resilience based on network topology and traffic measurements. By removing links or nodes to mimic deliberate attacks or random failures, topology robustness can then be evaluated through monitoring the performance of the chosen network measures (Zhang et al., 2015; Bhatia et al., 2015). An improved approach comes with the consideration of edge weights, which represent additional dimensions to the network topology such as travel time, cost, and travel distance (Calvert and Snelder, 2018). Also, the travel demand, route choice problems, and user equilibrium are taken into account as well. Proper examples include: a Network Robustness Index that developed based on the change of use equilibrium travel time under edge removal (Scott et al., 2006), a Vulnerability Index that incorporates traffic flow, travel time, capacities and availability of alternative routes (Murray-Tuite and Mahmassani, 2004), and a combined travel demand model that accounts for trip generation, destination, mode and route choices to measure the long-term equilibrium under single or multiple edge removal (Chen et al., 2007).

In systems engineering community, the resilience of a system is often perceived as a compound system capability, which consists of critical system functions and qualities (Boehm and Kukreja, 2017; Bruneau and Reinhorn, 2007). Systems perspective considers basic capabilities in system resilience. To differentiate with the network-based approaches, we term the resilience discussed under systems frameworks as the general resilience. As one of the urban critical infrastructure systems, some have attempted to study transportation resilience through system perspectives. For instance, Wang et al. (2015) proposed a day-to-day tolling scheme to promote the rapidity of road traffic resilience in external disruptions. Hosseini and Barker (2016) proposed a capability-based Bayesian network model (BNM) to measure system resilience in a case of waterway port systems. Boehm and Kukreja (2017) studied the system resilience as a compound capability and explored its ontology relationships with other system qualities.

The following gaps in the literature, including those reviewed above, have been identified here: (1) First, most existing studies are based on a single network measure or unilateral capability of the system effectiveness (Faturechi and Miller Hooks, 2014). In the road transportation context, there is a lack of integrated, system-level quantification based on a comprehensive picture of multi-facet enabling functions and system qualities. (2) Second, the long-term and dynamic features of general resilience and strength of the component qualities of a city's road transportation system has been rarely investigated and explored.

Therefore, we proposed a three-layer hierarchical BNM through a systems perspective, which based on multi-facet system functions and qualities, and evaluated the general resilience of Beijing's road system from 1997 to 2016. After ontologically defining the structure and conditional probability of the model, qualities were valued through multidimensional data, which was collected from various open sources, including annual statistics from central and local governments, transportation bureaus, and private companies. The aim is to fill the aforementioned gaps through answering three questions: (1) How can we model the general resilience of Beijing's road system with various qualities from a systems perspective? (2) What is the trend of the temporal change of its general resilience in the past two decades? (3) Which system qualities are the most critical factors that influence its overall resilience?

## METHODOLOGY

**Theory of Bayesian networks**

Bayesian networks are widely recognized as an effective and developed technique, based on Bayes' Theorem, for tackling probabilistic assessments and predictions with multiple variables. It is prominent for multi-criteria and multi-facet evaluation but with little application in system resilience modeling (Hosseini and Barker, 2016). Therefore, we applied this tool in our research and briefly introduce its basis here (readers can obtain detailed and in-depth knowledge through Jensen (1996) and Heckerman (1998)).

Mathematically, let V = {$X_1$, $X_2$, … $X_n$} be the set of variables in a BNM, where the conditional independence among variables are specified as the topology of the network. An outgoing edge from node $X_1$ to $X_2$ indicates that the probabilities of states in $X_2$ are dependent on the outcome in node $X_1$. We call $X_1$ is the parent node of $X_2$ and $X_2$ is the child node. In general, there are three types of nodes in a given BNM: (1) nodes without ingoing edges (no parent nodes), named root nodes, (2) nodes without outgoing edges (no child nodes) are labeled as leaf nodes, and (3) nodes with both ingoing and outgoing edges are called intermediate nodes. An illustrative example of a BNM is shown in Figure 1.

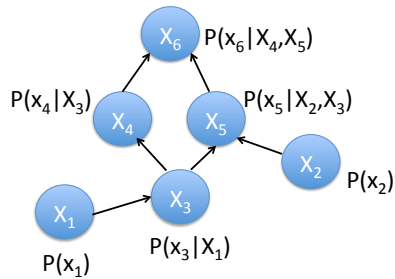

**Figure 1. An illustrative example of a BNM.**

The dependence or causal relationships among nodes are then measured by conditional probability. Each node in the network is associated with a Conditional

Probability Table (CPT) that defines the probability of the node states in the condition of its parent nodes. Therefore, we can calculate the joint probability for full BNM structure, by using the chain rule, into a factorized form with nodes' parents (Sun and Erath, 2015). In this way, for a network contains *n* variables, its full joint probability distribution can be expressed as:

$$P(X_1, X_2, \dots X_n) = \prod_{i=1}^{n}(P(X_i|\psi_i)) \tag{1}$$

where $P(X_1, X_2, \dots X_n)$ is the full joint probability distribution and $\psi_i$ is the parents of node $X_i$.

In principle, the structure and CPT of a BNM can be learned from a significant amount of observation data. However, these two are assessed by expert knowledge for models without full knowledge of observation data. Therefore, we used specialist knowledge and knowledge-based resilience frameworks to determine the structure and CPT for the proposed BNM.

**Elements and structure of the model**

For the structure of the BNM, we established a three-layer hierarchical topology. The first layer includes four resilience capability functions, the second layer consists of 10 system qualities related to the four functions, and the third layer is the indicators that act as elements of root nodes in the quality layer to kick off the calculation and probability propagation in the BNM.

The four resilience capability functions adapted were from a function-based resilience framework proposed by Heinimann and Hatfield (2017). In this work, they argued that biophysical functions - resistance, re-stabilization of critical functionality, and rebuilding and reconfiguration of that functionality - could be used to quantify system resilience as a promising start point. In urban road systems, these four functions depict four abilities of being resilient, that is, (1) To keep overall degradation within an acceptable limit; (2) To re-stabilize the performance after external disturbance; (3) To rebuild the performance after deteriorations; and (4) To reconfigure the system after recovery. These four elemental functions, in turn, outline the responsive actions of a road system from pre-event to post-event stages, which eventually contribute to its overall resilience.

Below the function layer, we identified 10 system qualities as the second layer (quality layer) of the network, which have significant contributions to those four functions and promote general system resilience. They include Availability, Serviceability, Robustness, Safety, Maintainability, Repairability, Affordability, Changeability, and Adaptability. For example, road systems have to maintain a suitable level of serviceability (ability to provide proper services) in order to be resilient in performance during a disruption event such as massive congestion. In the meantime, it only takes seconds to realize that we would expect a resilient road system to be safe against traffic accidents and economically affordable for all users. However, these qualities are not mutually independent. Here in Figure 2, we adopted the ontology structure (interdependence) of these 10 qualities (Boehm, 2016) in our BNM.

Next, one more indicator layer was added below the quality layer. These indicators were deliberately chosen from the road system performance. Here, we have 6 root quality nodes in the network (Serviceability, Robustness, Safety, Repairability, Affordability, and Adaptability). Each root node has a various number of indicators attached (Circular nodes in Figure 2) for calculation of the BNM. Serviceability includes the total road length, the average annual free-flow index, and the average annual passenger capacity; Robustness consists of the economic losses in natural disasters and the investment in disaster prevention in urban roads; Safety includes the number of injuries in traffic incidents and the number of accidents; Repairability contains the annual increasing rate of paved road area and the total number of personnel in the road transportation sector; Affordability includes the ratio of the expense on transportation, and Adaptability includes the number of granted transportation patents.

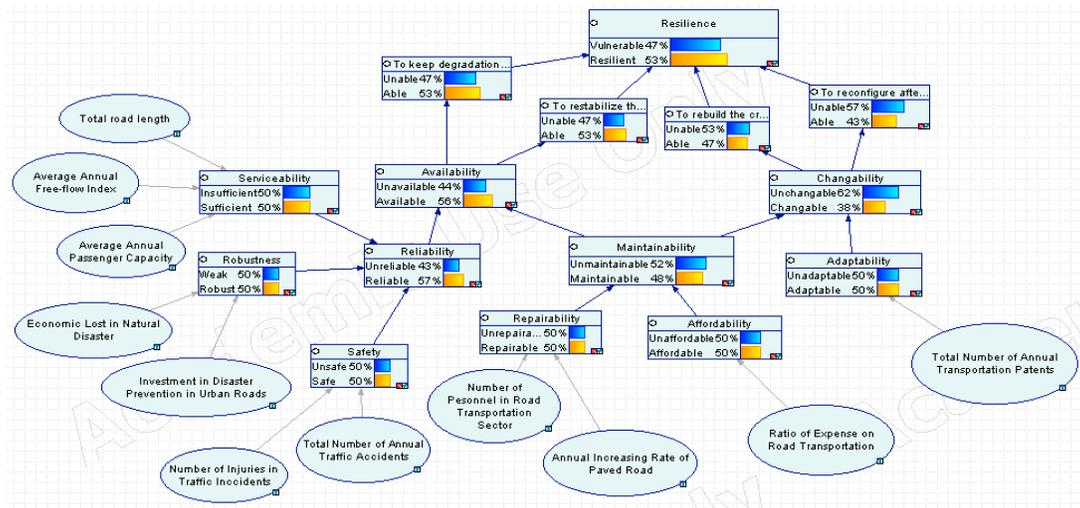

**Figure 2. The topology of the proposed BNM and determined CPT using GeNIe. For illustration, we also include indicators as circular nodes. (Here, we use technological innovation to represent system's ability to make changes, improvements, and adaptations).**

**Baseline of the proposed model**

In the proposed BNM, we set binary positive and negative states in all variable nodes for simplification. For example, node "Resilience" has "Resilient" and "Vulnerable" states and "Reliability" has "Reliable" and "Unreliable". By using open-source software - GeNIe, we obtained a baseline BNM with appropriate CPT input as Figure 2.

## EMPIRICAL STUDY

### Data collection and treatment

The data was collected from multiple sources, including the National Bureau of Statistics of China, the Ministry of Transport of the People's Republic of China, Tom-Tom Traffic Index, and Annual reports from Alibaba's Amap traffic group. We collected data for all the indicators from 1997 to 2016 in Beijing's urban area. The original dataset contains missing data and we treated them with the following methods: (1) For those indicators with no apparent growth rate and a small number of missing blanks, we fill in the blanks with the median value; And (2) For those with a linear or exponential growth, we fill in the missing data with regression predictions.

### Initial input in indicators

Because some qualities have multiple indicators, the overall values of these qualities were taken with an aggregated value. For clarification, we here demonstrate an example on how to obtain an aggregated value for a root quality node - Serviceability. We have three indicators contributing to the overall serviceability, namely total road length (million km), average annual free-flow index (dimensionless) and average annual passenger capacity (million people). Among them, the annual free-flow index can be calculated directly in a form of percentage as 2.06/2.5=82.4%, where 2.06 is the average annual congestion index of Beijing's traffic in 2016 and 2.5 is the maximum. Because congestion is a negative-effect indicator, we normalize it by 1-82.4%= 17.6% to represent the free-flow index. Values of the other two indicators were converted into probabilities with cumulative probability density (see Figure 3) with the mean and standard deviance. After obtaining probabilities of all three indicators, the yearly aggregated probability of the serviceability was taken as the algebraic mean of all probabilities contributed from these three indicators. Table 1 presents examples of how to obtain the aggregated probabilities of serviceability from raw data in the first five years. We can see that the overall serviceability of the road transportation system was not as high as expected in those observation years, especially the congestion has significantly deteriorated the overall level.

### General resilience of road system

The rest of the indicators were converted into probability density and aggregated to obtain values for other root nodes in a similar way. Table 2 presents the aggregated probabilities of all root nodes in the BNM during the observation years.

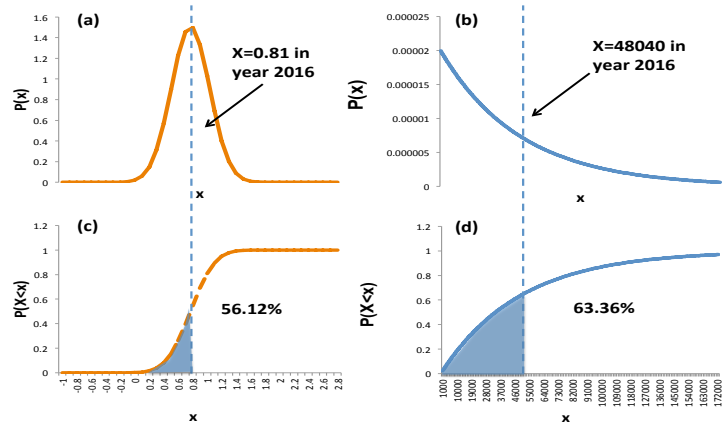

**Figure 3. Calculation of probabilities in two indicators. (a) is the assumed normal distribution of Total road length; (b) is the assumed exponential distribution of Average annual passenger capacity; (c) and (d) are the corresponding cumulative probability density values at given observation.**

**Table 1. The Procedure of data treatment for three indicators from serviceability quality in the first five years.**

| Year | Total road length (Mean = 0.77; STD = 0.27) | % | Average annual free-flow index | Average annual passenger capacity (Mean = 49166.30; STD = 50489.61) | % | Aggregated serviceability |
|---|---|---|---|---|---|---|
| 2016 | 0.81 | 56.12% | 17.6% | 48040 | 62.36% | 45.36% |
| 2015 | 0.81 | 56.12% | 17.5% | 49931 | 63.78% | 45.83% |
| 2014 | 0.81 | 56.12% | 18.1% | 52354 | 65.52% | 46.41% |
| 2013 | 0.79 | 53.13% | 18.3% | 52481 | 65.61% | 45.45% |
| 2012 | 0.79 | 53.13% | 17.9% | 132333 | 93.22% | 54.65% |
| … | … | … | … | … | … | … |

Figure 4 shows the quantification result of the resilience in 2016. As can be seen, the resilience of Beijing's road transportation system in 2016 had a moderate probability of 62% being resilient. The serviceability of the system was relatively unsatisfactory. One possible explanation, as mentioned above, could be that the well-known traffic congestion in Beijing severely deteriorates the overall serviceability. However, the performance of other qualities was mostly adequate with a higher chance of being positive. Particularly, the system's probability of being affordable was comparatively high (around 89%). This is true when considering the low cost of transportation in Beijing. In addition, we can also observe that the Maintainability, Adaptability, Reliability, and Availability of the system were all positively developed, and eventually led to a moderate level of resilience in 2016.

**Table 2: Probability of all quality root nodes in the proposed BNM.**

| Year | Service-ability | Robust-ness | Safety | Repairab-ility | Afford-ability | Adapt-ability |
|------|------|------|------|------|------|------|
| 2016 | 45.36% | 69.34% | 67.20% | 71.41% | 89.43% | 53.18% |
| 2015 | 45.83% | 95.80% | 70.17% | 74.96% | 89.43% | 60.16% |
| 2014 | 46.41% | 43.31% | 64.36% | 71.03% | 89.42% | 54.06% |
| 2013 | 45.45% | 87.54% | 64.75% | 73.88% | 89.42% | 50.81% |
| 2012 | 54.65% | 41.80% | 63.07% | 93.99% | 89.63% | 54.72% |
| 2011 | 46.81% | 37.49% | 56.59% | 41.76% | 89.30% | 52.45% |
| 2010 | 47.06% | 59.98% | 54.04% | 46.55% | 88.23% | 58.49% |
| 2009 | 45.92% | 46.57% | 57.32% | 46.34% | 89.65% | 45.63 |
| 2008 | 45.67% | 34.20% | 56.45% | 49.18% | 90.73% | 40.79% |
| 2007 | 18.39% | 28.86% | 46.01% | 25.64% | 89.41% | 47.20% |
| 2006 | 15.81% | 32.68% | 43.39% | 14.95% | 89.12% | 42.32% |
| 2005 | 40.67% | 28.22% | 41.27% | 61.24% | 88.99% | 44.75% |
| 2004 | 40.56% | 32.83% | 32.69% | 35.16% | 90.01% | 48.93% |
| 2003 | 34.83% | 43.31% | 25.34% | 34.35% | 87.84% | 48.51% |
| 2002 | 33.64% | 43.31% | 22.54% | 34.92% | 89.80% | 45.84% |
| 2001 | 31.13% | 43.31% | 38.76% | 33.80% | 89.14% | 51.31% |
| 2000 | 29.32% | 43.31% | 49.58% | 35.14% | 89.01% | 57.09% |
| 1999 | 27.64% | 43.31% | 10.40% | 35.14% | 88.72% | 75.48% |
| 1998 | 25.82% | 43.31% | 13.65% | 36.36% | 87.62% | 60.12% |
| 1997 | 24.73% | 43.31% | 23.28% | 40.51% | 87.62% | 52.70% |

Figure 5 shows the quantification results for all 20 years (the list shows the numerical values, and the figure is the bar plot with the approximated trend line). A clear "bathtub" or a "V-shape" trend line can be observed. Before 2006, the system had reasonable probabilities of being resilient with some unstable yearly fluctuations. However, there was a drastic drop in 2006. This instability could be a result of the rapid expansion, inadequate urban governance, and population growth during the early years. After 2006, the resilience had a noticeable increase possibly due to improvements in many aspects of Beijing's road system in the past decade.

**Sensitivity analysis and strength of influence**

Figure 6 presents the result of tests on sensitivity and strength of influence for each quality and functions. The thickness of the edges demonstrates the strength of influence. At the function level, the ability "to rebuild" had the most power to influence the overall resilience. At quality level, strong influence can be found in two parts. The first part was the route from Robustness-Reliability-Availability to the first two functions, and the second part was among Repairability via Maintainability, Changeability to the last two functions. We also observed a strong influence between Adaptability and Changeability.

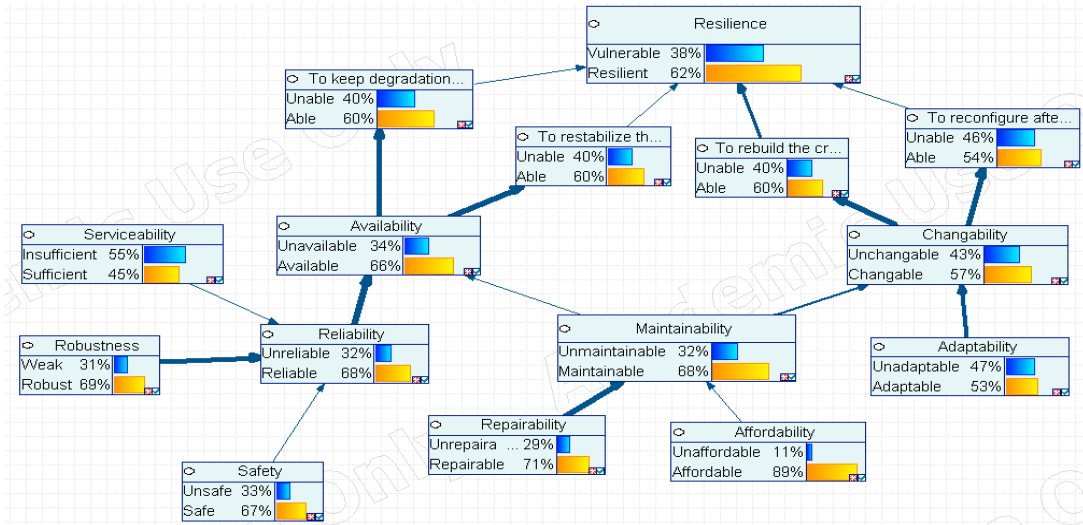

**Figure 4: Estimated general resilience of the road system in 2016.**

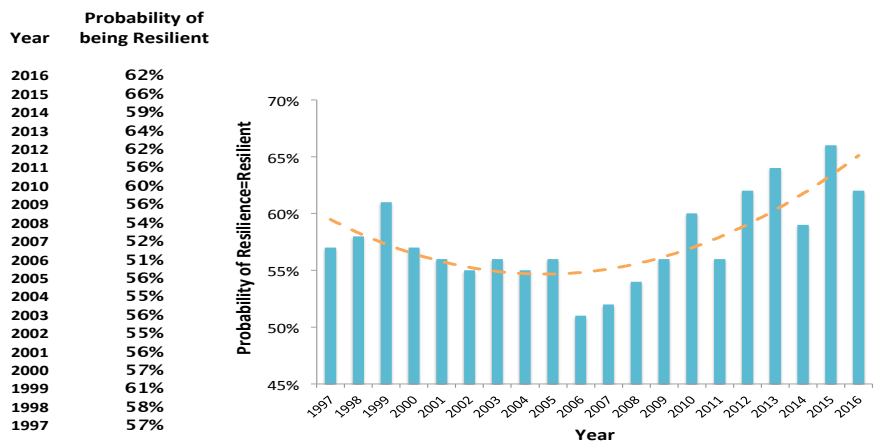

**Figure 5: Temporal result of resilience from 1997 to 2016.**

Node color represents the degree of sensitivity. The result shows that overall resilience of this transportation system was prominently sensitive to the changes in capability "to rebuild" at the function layer, and Robustness, Repairability, Changeability, and Adaptability at the quality layer. Further exploration of sensitivity confirmed this observation. By plotting the sensitivity tornado (Figure 7), we see that top 20 conditions (with "Resilience = Resilient") involve the Function of rebuilding, Adaptability, Changeability, Repairability, and Robustness. This intuitively indicates that the ability to rebuild its functionality performance after disruptions, its ability to quickly adapt and make changes, and its robustness and ability to repair damaged parts could be essential roles in the resilience of the road transportation system. Most importantly, effective strategies could be possibly developed based on this finding when building resilience into road systems.

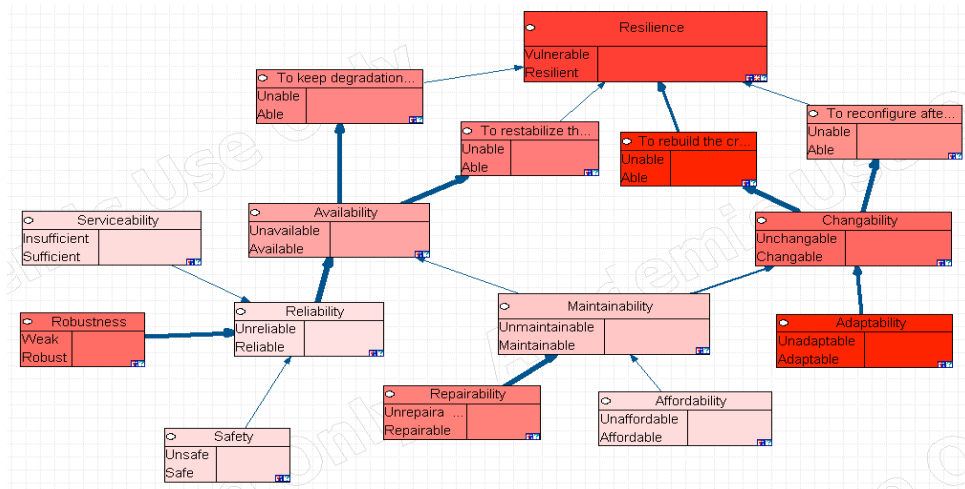

**Figure 6: Sensitivity test and strength of influence analysis.**

## CONCLUSIONS

This paper addressed the quantification issue of the general resilience in Beijing's road transportation system by proposing a Bayesian network model (BNM) with a system-level perspective. We have concluded the following:

- The proposed BNM is a promising tool for multi-dimensional and systematic analysis, instead of finding a one-size-fits-all quantification criterion for the resilience.
- For the past two decades (1997-2016), the general resilience of Beijing's road system exhibits a "V" shape in its trend, with the probability of being generally resilient between 50% and 70%, and at its minimum in 2006. There was a steep increase in such a probability since 2006.
- The analysis on sensitivity and strength of influence indicates that the general resilience of Beijing's road transportation system is most affected by its capabilities: (1) to rebuild its performance, (2) to be robust, (3) to adapt, (4) to change, and (5) to quickly repair damaged parts.

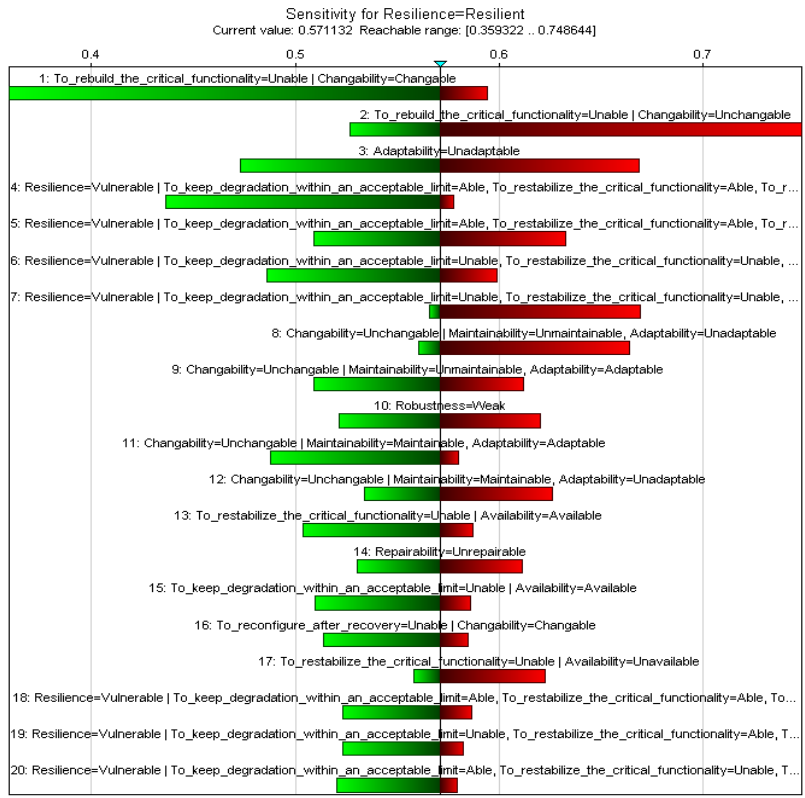

**Figure 7: Tornado plot of the sensitivity test**

**ACKNOWLEDGEMENT**

This research was conducted at the Future Resilient Systems at the Singapore-ETH Centre, which was established collaboratively between ETH Zurich and Singapore's National Research Foundation (FI 370074011).